\documentclass[pre,twocolumn,aps,showpacs]{revtex4}
\newcommand{\bec}[1]{\mbox{\boldmath $ #1$}}
\usepackage{graphicx}
\begin{document}
\bigskip
\bigskip
\title{Mean-field dynamo in a turbulence with shear and kinetic helicity fluctuations}
\author{Nathan Kleeorin}
\email{nat@bgu.ac.il}
\author{Igor Rogachevskii}
\email{gary@bgu.ac.il} \homepage{http://www.bgu.ac.il/~gary}
\affiliation{
 Department of Mechanical Engineering,
 The Ben-Gurion University of the Negev,
 POB 653, Beer-Sheva 84105, Israel}
\date{\today}
\begin{abstract}
We study effects of kinetic helicity fluctuations in a turbulence with large-scale shear using two different approaches: the spectral tau-approximation and the second order correlation approximation (or first-order smoothing approximation). These two approaches demonstrate that homogeneous kinetic helicity fluctuations alone with zero mean value in a sheared homogeneous turbulence cannot cause large-scale dynamo. Mean-field dynamo can be possible when kinetic helicity fluctuations are inhomogeneous which cause a nonzero mean alpha effect in a sheared turbulence. On the other hand, shear-current effect can generate large-scale magnetic field even in a homogeneous nonhelical turbulence with large-scale shear. This effect was investigated previously for large hydrodynamic and magnetic Reynolds numbers. In this study we examine the threshold required for the shear-current dynamo versus Reynolds number. We demonstrate that there is no need for a developed inertial range in order to maintain the shear-current dynamo (e.g., the threshold in the Reynolds number is of the order of 1).
\end{abstract}


\maketitle

\section{Introduction}

It has been widely recognized that astrophysical large-scale magnetic fields
originate due to the mean-field dynamo (see, e.g., \cite{M78,P79,KR80,ZRS83,RSS88,O03,BS05}). Such dynamo can be driven by the joint action of the mean kinetic helicity of turbulence and large-scale differential rotation. On the other hand, recently performed numerical experiments \cite{B05,YHS07,BRR07} have demonstrated existence of a nonhelical large-scale dynamo in a turbulence with a large-scale shear whereby mean $\alpha$ effect vanishes.
Note that a sheared turbulence is a universal feature in astrophysical \cite{O03,BS05,DCG05} and laboratory \cite{MP07} flows.

One of the possible mechanism of the nonhelical large-scale dynamo
in a homogeneous sheared turbulence is a shear-current effect that has been extensively studied during recent
years (see \cite{RK03,RK04,RKL06,RKCL06}). In particular, the
deformations of the original nonuniform magnetic field lines are
caused by upward and downward turbulent eddies. In a sheared
turbulence the inhomogeneity of the original mean magnetic field
breaks a symmetry between the influence of the upward and downward
turbulent eddies on the mean magnetic field. This creates the mean
electric current along the mean magnetic field and results in the nonhelical shear-current dynamo. Indeed, the large-scale velocity shear creates
anisotropy of turbulence that produces a contribution to the electromotive force,
${\bf W} {\bf \times} {\bf J}$, caused by the shear, where ${\bf W}$ is the
background large-scale vorticity due to the shear and  ${\bf J}$ is
the large-scale electric current. Joint effects of the electromotive force
${\bf W} {\bf \times} {\bf J}$ and stretching of the mean magnetic field due to the large-scale shear motions cause the mean-field dynamo instability. Note also that the numerical experiment with Taylor-Green forcing \cite{PON05} seems to be another example of a mean-field dynamo produced by a combined effect of a nonhelical turbulence and a complicated large-scale flow.

Other effect that might explain the nonhelical large-scale dynamo is related to  kinetic helicity fluctuations in a sheared turbulence. A problem associated with dynamics of large-scale magnetic field in the presence of kinetic helicity fluctuations in a shear-free turbulence has been formulated for the first time by Kraichnan \cite{K76} (see also \cite{M78}). In particular, he assumed that in small scales $l_\nu \ll l_{\rm turb} \ll l_0$ (and $\tau_\nu \ll \tau_{\rm turb} \ll \tau_0)$ there is a small-scale turbulence generated by forcing ${\bf F}^{(u)}$. On the other hand, in the scales $l_0 \ll l \ll l_\chi$ (and $\tau_0 \ll \tau \ll \tau_\chi$) there are kinetic helicity fluctuations (or $\tilde \alpha$ fluctuations) with a zero mean generated by forcing ${\bf F}^{(\chi)}$. The mean-field effects occur at large scales $L \gg l_\chi$ (and times $\tau_{_{L}} \gg \tau_\chi$), where the large-scale helicity is zero. The large-scale quantities are determined by double averaging over velocity fluctuations and over kinetic helicity fluctuations. It was found in \cite{K76} (see also \cite{M78}) that kinetic helicity fluctuations in a shear-free turbulence cause both, a reduction of turbulent magnetic diffusion and a large-scale drift velocity, ${\bf V}^{(\alpha)} \propto \bec{\nabla} \langle \tilde{\alpha}^2 \rangle$, of the mean magnetic field.

In recent time the dynamo problem related to kinetic helicity fluctuations has been extensively studied. In particular, various mathematical aspects of this problem has been discussed in \cite{SOK97} (see also \cite{F03}). Some numerical experiments which examine effects of kinetic helicity fluctuations have been performed in \cite{VB97,BRR07}. In particular, numerical simulations of the magnetic field evolution in accretion discs in \cite{VB97} have demonstrated that kinetic helicity fluctuations with a zero mean can result in generation of large-scale magnetic field.

It has been pointed out in \cite{SIL2000} that inhomogeneous kinetic helicity fluctuations in a sheared turbulence can produce a mean-field dynamo. In particular, a joint action of a large-scale shear and a nonzero mean alpha effect caused by the inhomogeneous kinetic helicity fluctuations can result in generation of a large-scale magnetic field, where the mean alpha effect in a sheared turbulence is proportional to ${\nabla} \langle \tilde{\alpha}^2 \rangle$. This mean-field dynamo is similar to the $\alpha\Omega$-dynamo. On the other hand, it has been suggested in \cite{PROC07} using phenomenological arguments that homogeneous kinetic helicity fluctuations in a homogeneous turbulence with shear may generate a large-scale magnetic field.

The main goal of this study is to examine a possibility for a nonhelical large-scale dynamo associated with homogeneous kinetic helicity fluctuations with a zero mean in a homogeneous turbulence with a large-scale shear. In a study of a sheared turbulence we use two different approaches, namely, the spectral tau-approximation \cite{RK04} and second order correlation approximation (SOCA, sometimes refers in astrophysical literature as first-order smoothing approximation (FOSA), see, e.g., \cite{RAS06,RUK06}). We also investigate dynamo effects associated with inhomogeneous kinetic helicity fluctuations in a sheared homogeneous turbulence.

This paper is organized as follows. In Sec.~II we formulate the
governing equations and outline the procedure of derivation based on the $\tau$-approach, that allows us to determine a contribution to the mean electromotive force caused by a combined action of the sheared turbulence and the kinetic helicity fluctuations. In Sec.~III we study the effects of kinetic helicity fluctuations in a sheared turbulence using the $\tau$-approach. In Sec.~IV we investigate similar effects using the SOCA-approach. In Sec.~V we discuss the threshold required for the shear-current dynamo versus Reynolds number. In Sec.~VI we draw concluding remarks. Finally, the detailed derivations of the effects of kinetic helicity fluctuations in a sheared turbulence using the $\tau$-approach and the SOCA-approach have been performed in Appendixes A and B, respectively.

\section{Governing equations and the $\tau$-approach}

In order to study mean-field dynamo in a turbulence
with kinetic helicity fluctuations and large-scale shear we use a procedure which is similar to that applied in \cite{RK04} for an investigation of a sheared turbulence. In particular, we use the following equations for fluctuations of velocity, ${\bf u}$, and magnetic field, ${\bf b}$, in order to determine the effect of shear on a turbulence:
\begin{eqnarray}
{\partial {\bf u} \over \partial t} &=& - ({\bf U} {\bf \cdot}
\bec{\nabla}) {\bf u} - ({\bf u} {\bf \cdot} \bec{\nabla}) {\bf U} -
\bec{\nabla} \biggl({p \over \rho}\biggr) + {1 \over \rho} \, \big[({\bf b} {\bf \cdot} \bec{\nabla}) {\bf B}
\nonumber\\
& & + ({\bf B} {\bf \cdot} \bec{\nabla}){\bf b}\big] + \nu \, \Delta {\bf u} + {\bf u}^{N} + {\bf F}^{(u)} +{\bf F}^{(\chi)} \;,
\label{A1} \\
{\partial {\bf b} \over \partial t} &=& ({\bf B} {\bf \cdot}
\bec{\nabla}){\bf u} - ({\bf u} {\bf \cdot} \bec{\nabla}) {\bf B} +
({\bf b} {\bf \cdot} \bec{\nabla}){\bf U} - ({\bf U} {\bf \cdot}
\bec{\nabla}) {\bf b}
\nonumber\\
& & + \eta \, \Delta {\bf b} + {\bf b}^N \; .
\label{A2}
\end{eqnarray}
The velocity field is assumed to be incompressible. Here ${\bf B} = \langle {\bf b}' \rangle$, $\, {\bf b}' = {\bf b} + {\bf B}$ is total magnetic field, the angular brackets $\langle ... \rangle$ denote averaging over ensemble of turbulent velocity field, the velocity ${\bf U} = \langle {\bf u}' \rangle = {\bf U}^{(S)} + {\bf V}$  includes an imposed large-scale sheared velocity ${\bf U}^{(S)}$, ${\bf u}' =  {\bf u} + {\bf U}$ is total velocity field, $\nu$  is the kinematic viscosity, $\eta$ is the magnetic diffusion due to electrical conductivity of the fluid, $\rho$ is the fluid density, $p$ are the fluctuations of total
(hydrodynamic and magnetic) pressure, the magnetic permeability
of the fluid is included in the definition of the magnetic field, ${\bf v}^{N}$ and ${\bf b}^{N}$ are the nonlinear terms, ${\bf F}^{(u)}$ and ${\bf F}^{(\chi)}$ are the stirring forces for velocity and kinetic helicity fluctuations, respectively.

Using Eqs.~(\ref{A1}) and (\ref{A2}) written in a Fourier space we
derive equations for the instantaneous two-point second-order
correlation functions of the velocity fluctuations $\langle u_i \,
u_j\rangle$, the magnetic fluctuations $\langle b_i \, b_j \rangle$,
and the cross-helicity tensor $\langle b_i \, u_j \rangle$. The
equations for these correlation functions are given by
Eqs.~(\ref{B6})-(\ref{B8}) in Appendix A. We split the tensor of
magnetic fluctuations into nonhelical, $h_{ij} = \langle b_i \, b_j
\rangle$, and helical, $h_{ij}^{(H)},$ parts. The helical part
$h_{ij}^{(H)}$ depends on the magnetic helicity, and it is
determined by the dynamic equation that follows from the magnetic
helicity conservation arguments (see, e.g.,
\cite{KR82,KRR94,GD94,KR99,KMRS2000,BF00,VC01,BB02}, and a review
\cite{BS05}).

The second-moment equations include the first-order spatial
differential operators $\hat{\cal N}$  applied to the third-order
moments $M^{(III)}$. A problem arises how to close the system, i.e.,
how to express the set of the third-order terms $\hat{\cal N}
M^{(III)}$ through the lower moments $M^{(II)}$ (see, e.g.,
\cite{O70,MY75,Mc90}). We use the spectral $\tau$-closure-approximation
which postulates that the deviations of the third-moment terms,
$\hat{\cal N} M^{(III)}({\bf k})$, from the contributions to these
terms afforded by the background turbulence, $\hat{\cal N}
M^{(III,0)}({\bf k})$, are expressed through the similar deviations
of the second moments, $M^{(II)}({\bf k}) - M^{(II,0)}({\bf k})$:
\begin{eqnarray}
\hat{\cal N} M^{(III)}({\bf k}) &-& \hat{\cal N} M^{(III,0)}({\bf
k})
\nonumber\\
&=& - {1 \over \tau_r(k)} \, \Big[M^{(II)}({\bf k}) - M^{(II,0)}({\bf k})\Big]
\;, \label{A3}
\end{eqnarray}
(see \cite{O70,PFL76,KRR90,KMR96,RK04}), where $\tau_r(k)$ is the
scale-dependent relaxation time, which can be identified with the
correlation time of the turbulent velocity field for large
hydrodynamic and magnetic Reynolds numbers. The quantities
with the superscript $(0)$ correspond to the background shear-free
turbulence with a zero mean magnetic field. We apply the
spectral $\tau$ approximation only for the nonhelical part $h_{ij}$
of the tensor of magnetic fluctuations. Note that a justification of
the $\tau$ approximation for different situations has been performed
in numerical simulations and analytical studies in
\cite{BS05,BF02,FB02,BK04,BSM05,SSB07,BS07} (see also detailed
discussion in \cite{RK07}, Sec. 6).

We assume that the characteristic time of variation of the
magnetic field ${\bf B}$ is substantially larger than the
correlation time $\tau(k)$ for all turbulence scales. This allows us
to get a stationary solution for the equations for the second-order
moments, $M^{(II)}$. We split all second-order correlation
functions, $M^{(II)}$, into symmetric $h_{ij}^{(s)} = [h_{ij}({\bf
k}) + h_{ij}(-{\bf k})] / 2$ and antisymmetric $h_{ij}^{(a)} =
[h_{ij}({\bf k}) - h_{ij}(-{\bf k})] / 2$ parts with respect to the
wave vector ${\bf k}$. For the integration in ${\bf k}$-space we
have to specify a model for the background shear-free turbulence.
A non-helical part of the homogeneous background turbulence is
given by the following equations
\begin{eqnarray}
\langle u_i \, u_j \rangle^{(0)}({\bf k}) &=& \langle {\bf u}^2
\rangle \, \Big(\delta_{ij} - {k_i k_j \over k^2} \Big) \, {E(k)
\over 8 \pi k^{2}} \;,
\label{A4} \\
\langle b_i \, b_j \rangle^{(0)}({\bf k}) &=& \langle {\bf b}^2
\rangle \, \Big(\delta_{ij} - {k_i k_j \over k^2} \Big)  \,
{E(k) \over 8 \pi k^{2}} \;,
\label{A5}
\end{eqnarray}
where $\delta_{ij}$ is the Kronecker tensor, the energy spectrum function is $E(k) = k_0^{-1} \, (q-1) \, (k / k_{0})^{-q}$, the wave number $k_{0} = 1 / l_{0}$, the length $l_{0}$ is the maximum scale of turbulent motions. The turbulent correlation time is $\tau(k) = C \, \tau_0 \, (k / k_{0})^{-\mu}$, where the coefficient $C(q,\mu)=(q-1+\mu)/(q-1)$. This value of the coefficient $C$ corresponds to the standard form of the turbulent diffusion coefficient in the isotropic case, i.e.,
$\eta_{_{T}} = \int \tau(k) \, [\langle {\bf u}^2 \rangle \, E(k)]
\, dk = \tau_0 \, \langle {\bf u}^2 \rangle /3$. Here the time
$\tau_0 = l_{0} / \sqrt{\langle {\bf u}^2 \rangle}$ and
$\sqrt{\langle {\bf u}^2 \rangle}$ is the characteristic turbulent
velocity in the scale $l_{0}$. For the Kolmogorov's type background
turbulence (i.e., for a turbulence with a constant energy flux over
the spectrum), the exponent $\mu=q-1$ and the coefficient $C=2$.
In the case of a turbulence with a scale-independent correlation time,
the exponent $\mu=0$ and the coefficient $C=1$.

On the other hand, a helical part of the background turbulent velocity
field is given by the following equation:
\begin{eqnarray}
\langle u_i \, u_j \rangle^{(0, \chi)}({\bf k}) &=& i \, \chi^v \varepsilon_{jin} \, k_n \, {E_\chi(k) \over 8 \pi k^{4}} \;,
\label{AA4}
\end{eqnarray}
where $\varepsilon_{ijk}$ is the fully antisymmetric Levi-Civita tensor, $\chi^v = \langle {\bf u} {\bf \cdot} (\bec{\nabla} \times {\bf u})
\rangle$ is the kinetic helicity, the spectrum function is
$E_\chi(k) = k_0^{-1} \, C_\chi \, (k / k_{0})^{- q}$ and $ C_\chi = q-1$ for large hydrodynamic Reynolds numbers. In the scales
$l_0 \ll l \ll l_\chi$ there are fluctuations of kinetic helicity  $\chi^v$ (see Sec.~3).

Using the solution of the derived second-moment equations, we determine the contributions to the electromotive force, ${\cal
E}_{i}^{(S,\alpha)} = \varepsilon_{imn} \, \int \langle b_n \, u_m
\rangle_{\bf k}^{(S,\alpha)} \,d {\bf k}$, caused by a combined action of the sheared turbulence and the kinetic helicity fluctuations (see Appendix A).

\section{Effects of kinetic helicity fluctuations in a sheared turbulence: $\tau$-approach}

The procedure described in Sec.~II allows us to determine
the contributions to the electromotive force (in particular, to the $\bec{\hat\alpha}$ tensor) caused by a combined action of the sheared turbulence and the kinetic helicity fluctuations (for details see Appendix A). This procedure
yields the equation for the evolution of the magnetic field ${\bf B}$:
\begin{eqnarray}
{\partial {\bf B} \over \partial t} = \bec{\nabla} {\bf \times} \Big(
\bec{\hat\alpha} {\bf B} + {\bf U}^{(S)} {\bf \times} {\bf B} - \eta_{_{T}} {\bf J}\Big) + {\bf B}^N\;,
\label{B5}
\end{eqnarray}
where $\eta_{_{T}}$ is the turbulent magnetic diffusion coefficient, ${\bf J}= \bec{\nabla} {\bf \times} {\bf B}$ is the electric current, ${\bf B}^N$ are the nonlinear terms, ${\bf U}^{(S)}$ is the imposed background sheared velocity and we assume for simplicity that ${\bf V}=0$. Here the total $\bec{\hat\alpha}$ tensor is given by
\begin{eqnarray}
\alpha_{ij} = \tilde \alpha \, \delta_{ij} + \alpha_{ij}^{(S,\alpha)} \;,
\label{B4}
\end{eqnarray}
where $\tilde \alpha \, \delta_{ij}$ determines a contribution to the total $\bec{\hat\alpha}$ tensor caused by a shear-free turbulence, while $\alpha_{ij}^{(S,\alpha)}$ describes a contribution to the $\bec{\hat\alpha}$ tensor caused by a combined action of the sheared turbulence and the kinetic helicity fluctuations. The tensor $\alpha_{ij}^{(S,\alpha)}$ reads
\begin{eqnarray}
\alpha_{ij}^{(S,\alpha)} = - \tilde \alpha \, \tau_0 \, \Big[C_1 \, (\partial U)_{ij}^{(S)} + C_2 \, \varepsilon_{ijn} \, W_n^{(S)} \Big] \;,
\label{B3}
\end{eqnarray}
where  $(\partial U)_{ij}^{(S)} = (\nabla_i U_{j}^{(S)} + \nabla_j U_{i}^{(S)}) / 2$, $\, {\bf W}^{(S)}= \bec{\nabla} {\bf \times} {\bf U}^{(S)}$, the coefficients $C_1 = (3 I  / 5) \, (3- 2 \mu)$, $\, C_2 = I  / 2$, and the parameter $I$ is given by
\begin{eqnarray}
I = \tau_0^{-2} \, \int \tau^2(k) \, E(k) \, dk = {(q-1+\mu)^2 \over (q-1+2 \mu) \, (q-1)} \; .
 \label{SB12}
\end{eqnarray}
For the Kolmogorov's type background turbulence (i.e., for a turbulence with a constant energy flux over the spectrum), the exponent $\mu=q-1$ and the coefficients $C_1 = (4  / 5) \, (3- 2 \mu)$, $\,C_2 = 2  / 3$.
The tensor $\alpha_{ij}^{(S,\alpha)}$ has been derived in Appendix A.

Using Eq.~(\ref{B5}) we derive equation for the correlation function $\langle \alpha_{ij} B_p \rangle^{(\alpha)}$:
\begin{eqnarray}
&&{\partial \over \partial t} \langle \alpha_{ij} B_p \rangle^{(\alpha)} = \varepsilon_{pmn} \, \Big[(\nabla_m \bar B_k) \, \langle \alpha_{ij}  \alpha_{nk} \rangle^{(\alpha)}
\nonumber\\
&&  \quad \quad\quad + \; \bar B_k \, \langle \alpha_{ij} \nabla_m \alpha_{nk} \rangle^{(\alpha)} \Big] + \langle \alpha_{ij} B_n \rangle^{(\alpha)} \,\nabla_n U^{(S)}_p
\nonumber\\
&& \quad \quad\quad + \; \hat{\cal N} \langle \alpha_{ij} B_p \rangle^{(\alpha)} \;,\label{B9}
\end{eqnarray}
where $\bar{\bf B}=\langle {\bf B} \rangle^{(\alpha)}$, the brackets $\langle ... \rangle^{(\alpha)}$ denote an averaging over random $\tilde \alpha$ fluctuations, $\bar {\bf J}= \bec{\nabla} {\bf \times} \bar {\bf B}$ and $\hat{\cal N} \langle \alpha_{ij} B_p \rangle^{(\alpha)}$ determines the third-order moments caused by the nonlinear terms, which include also the turbulent diffusion term. In Eq.~(\ref{B9}) we use the spectral $\tau$ approximation~(\ref{A3}), whereby the relaxation time is of the order of the time $\tau_\chi$. We also take into account that the characteristic time of variation of the mean magnetic field $\bar{\bf B}$ is substantially larger than the relaxation time $\tau_\chi$. Then the steady state solution of Eq.~(\ref{B9}) allows us to determine the correlation function $\langle \alpha_{ij} B_j \rangle^{(\alpha)}$, that is given by Eq.~(\ref{B10}) in Appendix A.

Now let us consider for simplicity a linear mean velocity shear ${\bf U}^{(S)} = (0, Sx, 0)$ and $ {\bf W}^{(S)} = (0,0,S)$ with $S \, \tau_0 \ll 1$. We also consider the mean magnetic field $\bar{\bf B}$ in a most simple form $\bar{\bf B} = (\bar B_x(z), \bar B_y(z), 0)$. Therefore, the correlation function $\langle \alpha_{yj} B_j \rangle^{(\alpha)}$ is given by
\begin{eqnarray}
\langle \alpha_{yj} B_j \rangle^{(\alpha)} &=& {1 \over 2} \, \tau_\chi \, \Big[ S \, \big(\tau_\chi + 2C_2 \, \tau_0 \big) \, \big[2 \bar J_x - (\bar{\bf B} {\bf \times}\bec{\nabla})_x \big]
\nonumber\\
&&  + 2 \bar J_y - (\bar{\bf B} {\bf \times}\bec{\nabla})_y \Big] \, \langle \tilde \alpha^2 \rangle^{(\alpha)} \;,
 \label{B11}
\end{eqnarray}
where we used Eq.~(\ref{B10}) given in Appendix A.

\subsection{Inhomogeneous kinetic helicity fluctuations}

Let us first analyze inhomogeneous kinetic helicity fluctuations.
The last term in Eq.~(\ref{B11}) describes a large-scale drift velocity of the mean magnetic field:
\begin{eqnarray}
{\bf V}^{(\alpha)} = {\tau_\chi \over 2} \, \bec{\nabla} \langle \tilde{\alpha}^2 \rangle^{(\alpha)} \;,
\label{AB1}
\end{eqnarray}
where $\tau_\chi = l_\chi^2/\eta_{_{T}}$. The third term in Eq.~(\ref{B11}) determines a negative contribution to the turbulent magnetic diffusion of the mean magnetic field:
\begin{eqnarray}
\eta^{(\alpha)}_{_{T}} = - \tau_\chi \, \langle \tilde{\alpha}^2 \rangle^{(\alpha)}
\; .
\label{AB2}
\end{eqnarray}
Note that the total turbulent magnetic diffusion coefficient, $\eta_{_{T}} + \eta^{(\alpha)}_{_{T}}$, should be positive. The reduction of the turbulent magnetic diffusion and the large-scale drift velocity, ${\bf V}^{(\alpha)} \propto \bec{\nabla} \langle \tilde{\alpha}^2 \rangle$, of the mean magnetic field caused by kinetic helicity fluctuations have been obtained previously in \cite{K76} (see also \cite{M78}) using the SOCA approach. The second term in Eq.~(\ref{B11}) describes a mean $\alpha$ effect:
\begin{eqnarray}
\bar{\alpha}^{(S,\alpha)} = -{\tau_\chi \, S \over 2} \, \big(\tau_\chi + 2C_2 \, \tau_0 \big) \, {\nabla}_z \langle \tilde{\alpha}^2 \rangle^{(\alpha)} \;,
\label{AB3}
\end{eqnarray}
caused by a combined action of a large-scale shear and inhomogeneous kinetic helicity fluctuations. This effect can result in a mean-field dynamo (see \cite{SIL2000}) that acts as the $\alpha\Omega$-dynamo. The first term in Eq.~(\ref{B11}), that is proportional to $\bar J_x$, contributes to the coefficient $\sigma_\alpha$ determined by Eq.~(\ref{B15}) below.

\subsection{Homogeneous kinetic helicity fluctuations}

Now let us consider homogeneous kinetic helicity fluctuations $(\bec{\nabla} \langle \tilde{\alpha}^2 \rangle^{(\alpha)}=0)$ when the mean $\alpha$ effect vanishes. The total contribution to the mean electromotive force caused by the sheared turbulence and the kinetic helicity fluctuations is ${\cal E}_{i}^{(S,\alpha)} = \langle \alpha_{ij} B_j \rangle^{(\alpha)} + b_{ijk}^{(S)} \, \nabla_k \, \bar B_{j}$.
Here the tensor $b_{ijk}^{(S)}$ determines the shear-current effect and is given by
\begin{eqnarray}
b_{ijk}^{(S)} = l_0^2 \, [C_3 \, \varepsilon_{ikn} \, (\partial U)_{nj}^{(S)} + C_4 \,  \delta_{ij} \, W_k^{(S)}] \;,
 \label{B12}
\end{eqnarray}
(see \cite{RK04}), where $C_3 = I \, [1 - 2 \, \mu + \epsilon \, (9 + 10 \, \mu)] /30$, $\, C_4 = I \, [3 - 2 \, \mu - \epsilon \, (5 + 2 \, \mu)]/60 $, the parameter $\epsilon = E_m / E_v$, $\,\, E_m$ and $E_v$ are the magnetic and kinetic energies per unit mass in the background turbulence, and $I$ is determined by Eq.~~(\ref{SB12}).
Magnetic fluctuations in the background turbulence are caused by a small-scale dynamo (see, e.g., \cite{ZRS90,RK97,SIC07}).

The $y$-component of the mean electromotive force caused by the sheared turbulence and the kinetic helicity fluctuations reads
\begin{eqnarray}
{\cal E}_{y}^{(S,\alpha)} = l_0^2 \, S \, (\nabla_z \bar B_y) \, \sigma_{_{S}} - \eta^{(\alpha)}_{_{T}} \, \bar J_y \;,
 \label{B13}
\end{eqnarray}
where
\begin{eqnarray}
\sigma_{_{S}} &=& \sigma_{_{B}} - {\tau_\chi^2  \over \tau_0^2} \, {\langle \tilde \alpha^2 \rangle^{(\alpha)} \over u_0^2} \, \sigma_\alpha \;,
 \label{BB13}\\
\sigma_{_{B}} &=& {1 \over 2} \,C_3 + C_4 = {I \over 15}
\, [1 - \mu + \epsilon \, (1 + 2 \, \mu)] \;,
 \label{B14}\\
\sigma_\alpha &=& 1 + 2C_2 \, {\tau_0  \over \tau_\chi} = 1 + \, {4 \, \tau_0  \over 3 \, \tau_\chi} ,
\nonumber\\
 \label{B15}
\end{eqnarray}
and $u_0 = l_0 / \tau_0$. The parameter $\sigma_{_{B}}$ describes the shear-current effect, while the parameter $\sigma_\alpha$ determines the combined effect of the kinetic helicity fluctuations and the sheared turbulence. Equation~(\ref{B14}) for the parameter $\sigma_{_{B}}$ has been derived in \cite{RK04} for the case of large hydrodynamic and magnetic Reynolds numbers. In Sec.~V we determine the parameter $\sigma_{_{B}}$ for the case when hydrodynamic and magnetic Reynolds numbers are not large.

\subsection{Mean-field dynamo}

The equation for the evolution of the mean magnetic field, $\bar{\bf B} = (\bar B_x(z), \bar B_y(z), 0)$, reads
\begin{eqnarray}
{\partial \bar B_x \over \partial t} &=& - \sigma_{_{S}} \, S \, l_0^2 \,
\bar B''_y + \tilde \eta_{_{T}} \, \bar B''_x  \;,
 \label{E2}\\
{\partial \bar B_y \over \partial t} &=& S \, \bar B_x + \tilde \eta_{_{T}} \, \bar B''_y
\;,
 \label{E3}
\end{eqnarray}
where $\tilde \eta_{_{T}} = \eta_{_{T}} + \eta^{(\alpha)}_{_{T}}$ and $\bar B''_i = \partial^2 \bar B_i / \partial z^2 $. Here we neglect
small contributions to the coefficient of turbulent magnetic
diffusion caused by the shear motions because $S \tau_0 \ll 1$.
The solution of equations~(\ref{E2}) and~(\ref{E3}) we seek for in
the form $ \propto \exp(\gamma_{_{S}} \, t + i K_z \, z) ,$ where the growth rate, $\gamma_{_{S}}$, of the mean magnetic field is given by
\begin{eqnarray}
\gamma_{_{S}} = S \, l_0 \, \sqrt{\sigma_{_{S}}} \, K_z -
\tilde \eta_{_{T}} \, K_z^2 \; .
 \label{E5}
\end{eqnarray}
The necessary condition for the magnetic dynamo instability is
$\sigma_{_{S}} > 0$. The parameter $\sigma_\alpha > 0$ (see Eqs.~(\ref{BB13}) and~(\ref{B15})). This implies that homogeneous kinetic helicity fluctuations in a homogeneous turbulence with shear cause a negative contribution to the parameter $\sigma_{_{S}}$. Therefore, homogeneous kinetic helicity fluctuations in a sheared turbulence act against mean-field dynamo (see Eqs.~(\ref{BB13}) and~(\ref{E5})), while the shear-current effect may cause the generation of the large-scale magnetic field when the parameter $\sigma_{_{B}} > 0$ (see also Sec.~V).

Note that two effects determined by the parameters $\sigma_{_{B}}$ and $\sigma_\alpha$, can be interpreted as the off-diagonal terms in the tensor of turbulent magnetic diffusion. The kinetic helicity fluctuations cause a negative contribution to the diagonal components $\propto \eta^{(\alpha)}_{_{T}}$ of turbulent magnetic diffusion of the mean magnetic field (see Eq.~(\ref{AB2})). On the other hand, the kinetic helicity fluctuations also results in a negative contribution to the off-diagonal term $\propto \sigma_{_{S}} \propto - \sigma_\alpha$ in the tensor of turbulent magnetic diffusion.

In order to determine the threshold required for the excitation of
the mean-field dynamo instability, we consider the solution of
Eqs.~(\ref{E2}) and~(\ref{E3}) with the following boundary
conditions: $\bar {\bf B}(t,|z|=L) = 0$ for a layer of the thickness $2L$ in the $z$ direction. The solution for the mean magnetic field is
determined by
\begin{eqnarray}
\bar B_y(t,z) &=& B_0 \, \exp(\gamma_{_{S}} \, t) \, \cos (K_z \, z + \varphi) \;,
\label{M5} \\
\bar B_x(t,z) &=& l_0 \, K_z \, \sqrt{\sigma_{_{S}}} \, B_y(t,z) \; .
\label{M6}
\end{eqnarray}
For the symmetric mode the angle $\varphi =\pi \, n$ and the
large-scale wave number $K_z=(\pi / 2) (2m + 1)\, L^{-1}$, where $n, m = 0, 1, 2, ... \,$. For this mode the mean magnetic field is
symmetric relative to the middle plane $z=0$. Let us introduce the
dynamo number $D= (l_0 \, S_\ast / L)^2 \, \sigma_{_{S}}$, where
parameter $S_\ast = S \, L^2 / \tilde \eta_{_{T}}$ is the dimensionless
shear number. For the symmetric mode the mean magnetic field is
generated due to the shear-current effect when the dynamo number $D
> D_{\rm cr} = (\pi^2/4) (2m + 1)^2$. For the antisymmetric mode the
angle $\varphi =(\pi / 2) \, (2n+1)$ with $n = 0, 1, 2, ...$, the
wave number $K_z=\pi \, m \, L^{-1}$ and the magnetic field is
generated when the dynamo number $D > D_{\rm cr} = \pi^2 \, m^2$,
where $m = 1, 2, 3, ... \,$. The maximum growth rate of the mean
magnetic field, $\gamma_{\rm max} = S^2 \, l_0^2 \, \sigma_{_{S}} / 4 \tilde \eta_{_{T}}$, is attained at $K_z = S \, l_0 \, \sqrt{\sigma_{_{S}}} / 2 \tilde \eta_{_{T}}$. Therefore, the characteristic scale of the mean magnetic field variations $L_B = 2 \pi  / K_z = 4 \, u_0 / (S \, \sqrt{\sigma_{_{S}}})$.

\section{Effects of kinetic helicity fluctuations in a sheared turbulence: the SOCA-approach}

Now we study the effects of kinetic helicity fluctuations in a
sheared turbulence using a second order correlation approximation (SOCA). This approximation is valid only for small hydrodynamic Reynolds numbers. Even in a highly conductivity limit (large magnetic Reynolds numbers), SOCA can be valid only for small Strouhal numbers (i.e., for
very short correlation time).

The procedure of the derivation of the electromotive force in a
homogeneous turbulence with shear and kinetic helicity fluctuations
is as follows (for details see Appendix B). We use Eqs.~(\ref{A1})
and ~(\ref{A2}) for fluctuations of velocity and magnetic fields,
exclude the pressure term from the equation of motion~(\ref{A1}) by
calculation $\bec{\nabla} {\bf \times} (\bec{\nabla} {\bf \times} {\bf
u})$. We rewrite the obtained equation and Eq.~(\ref{A2})
in a Fourier space and apply the two-scale approach (i.e., we use
large-scale and small-scale variables).
We neglect nonlinear terms but keep molecular dissipative terms
in Eqs.~(\ref{A1}) and ~(\ref{A2}) for fluctuations of velocity
and magnetic fields. We seek for a solution for fluctuations of
velocity and magnetic fields as an expansion for weak velocity shear.

This procedure allows us to determine the contributions to the
electromotive force caused by a combined action of the sheared
turbulence and the kinetic helicity fluctuations. In particular,
the tensor $\alpha_{ij}^{(S,\alpha)}$ caused by the kinetic helicity fluctuations in a sheared turbulence is given by
\begin{eqnarray}
\alpha_{ij}^{(S,\alpha)} = - \tilde \alpha \, \tau_0 \, \Big[\tilde C_1 \, (\partial U)_{ij}^{(S)} + \tilde C_2 \, \varepsilon_{ijn} \, W_n^{(S)} \Big] \;,
\label{BB3}
\end{eqnarray}
(for details see Appendix B), where the coefficients $\tilde C_1$ and $\tilde C_2$ are
\begin{eqnarray}
\tilde C_1 &=& - {C_\ast \over 10} \,  (11\, {\rm Re} + 3 {\rm Rm})  \;, \quad
\tilde C_2  = {C_\ast \over 8} \, ({\rm Rm} -2 {\rm Re}) \;,
\nonumber\\
\label{C9}\\
C_\ast &=& {q+1 \over q+3} \, \biggl[{1 - \big(l_\ast / l_0\big)^{q+3} \over 1 - \big(l_\ast / l_0\big)^{q+1}} \biggr] \;,
\label{C10}
\end{eqnarray}
and ${\rm Re} = u_0 \, l_0 / \nu$ is the hydrodynamic Reynolds number, ${\rm Rm} = u_0 \, l_0 / \eta$ is the magnetic Reynolds number. Here we take into account that
\begin{eqnarray}
\tilde \alpha_{ij} \equiv i \, \varepsilon_{inm} \, \int k_j \, G_\eta \, f_{mn}^{(0,\chi)} \, d{\bf k} \, d\omega = \tilde \alpha \, \delta_{ij} \;,
\label{C11}
\end{eqnarray}
where
\begin{eqnarray}
\tilde \alpha &=& - {C_0 \over 3} \, \tau_0 \,  \chi^v \, {\rm Rm} \;,
\label{C12}\\
C_0 &=&  {\tilde C_\chi \over q+1} \, \biggl[1 - \Big({l_\ast \over l_0}\Big)^{q+1} \biggr] \;,
\nonumber
\end{eqnarray}
the function $G_\eta(k, \omega) = (\eta k^2 - i \omega)^{-1}$ and $f_{ij}^{(0,\chi)}({\bf k},\omega) \equiv \langle u_i \, u_j \rangle^{(0, \chi)}({\bf k},\omega) = - i \, \chi^v \, \varepsilon_{ijn} \, k_n \, \tilde E_\chi(k, \omega) /(8 \pi k^4)$. To integrate in ${\bf k}$ and $\omega$ space we used the following model for the spectrum function $\tilde E_\chi(k, \omega) = \tilde C_\chi \, k_0^{-1} \,  (k / k_0)^{- q} \, \delta(\omega)$, where $\tilde C_\chi = (q-1) \, \big[1 - \big(l_\ast / l_0\big)^{q-1} \big]^{-1}$ and the wave number $k$ varies in the interval from $l_0^{-1}$ to $l_\ast^{-1}$. Since SOCA is valid for small hydrodynamic Reynolds numbers the scale $l_\ast$ is not related to Kolmogorov (viscous) scale $l_\nu$.

In the scales $l_0 \ll l \ll l_\chi$ there are fluctuations of kinetic helicity  $\chi^v$ (or fluctuations of $\tilde \alpha$). In order to determine the correlation function $\langle \alpha_{ij} B_p \rangle^{(\alpha)}$ we use Eq.~(\ref{B5}) in which $\eta_{_{T}}$ is replaced by $\eta + \eta_{_{T}}$ and we neglect the nonlinear terms.
Solving this equation in a Fourier space we determine the magnetic field $B_i(\tilde{\bf K}, \Omega)$, where the wave vector $\tilde{\bf K}$ and the frequency $\Omega$ are in the spatial scales $l_0 \ll l \ll l_\chi$ and the time scales $\tau_0 \ll \tau \ll \tau_\chi$. Multiplying the magnetic field $B_j(\tilde{\bf K}, \Omega)$ by the tensor $\alpha_{ij}= \tilde \alpha \, \delta_{ij} + \alpha_{ij}^{(S,\alpha)} $ and averaging over kinetic helicity fluctuations we determine the correlation function $\langle \alpha_{ij} B_j \rangle^{(\alpha)}$. It is given by Eq.~(\ref{C13}) in Appendix B.

Now we consider a linear mean velocity shear ${\bf U}^{(S)} = (0,
Sx, 0)$ and assume that the mean magnetic field $\bar{\bf B}$ has the form $\bar{\bf B} = (\bar B_x(z), \bar B_y(z), 0)$. Therefore, Eq.~(\ref{C13}) yields the correlation function $\langle \alpha_{yj} B_j \rangle^{(\alpha)}$:
\begin{eqnarray}
&&\langle \alpha_{yj} B_j \rangle^{(\alpha)} = {1 \over 2} \, \int G_T \, \Big[ S \, \big(G_T + 2\tilde C_2 \, \tau_0\big) \, \big[2 \bar J_x \nonumber\\
&& \quad \quad - (\bar{\bf B} {\bf \times}\bec{\nabla})_x \big] + 2 \bar J_y - (\bar{\bf B} {\bf \times}\bec{\nabla})_y \Big] \, \langle \tilde \alpha^2 \rangle^{(\alpha)}_{\tilde K} \, d\tilde K \, d\Omega  \;,
\nonumber\\
 \label{C14}
\end{eqnarray}
where $G_T(\tilde K, \Omega) = [(\eta + \eta_{_{T}}) \tilde K^2 - i \Omega]^{-1}$.

\subsection{Homogeneous kinetic helicity fluctuations}

Let us consider homogeneous kinetic helicity fluctuations. The $y$-component of the mean electromotive force caused by the sheared turbulence and the kinetic helicity fluctuations is given by Eq.~(\ref{B13}), the parameter $\sigma_{_{S}}$ is determined by Eq.~(\ref{BB13}) whereby the time $\tau_\chi$ is replaced by the time $\tilde \tau_\chi = l_\chi^2/(\eta + \eta_{_{T}})$, and the parameter $\sigma_\alpha$ is given by
\begin{eqnarray}
\sigma_\alpha &=& {1 \over \tilde \tau_\chi^2 \, \langle \tilde \alpha^2 \rangle^{(\alpha)}} \, \int G_T \, \big(G_T + 2\tilde C_2 \, \tau_0\big) \, \langle \tilde \alpha^2 \rangle^{(\alpha)}_{\tilde K} \, d\tilde K \, d\Omega
\nonumber\\
&=& {\tilde q-1 \over \tilde q+3}  + {C_\ast \, \tilde C_0 \over 4} \, \Big({\tau_0  \over \tilde \tau_\chi}\Big) \, ({\rm Rm} -2 {\rm Re})   \; .
 \label{BB15}
\end{eqnarray}
Here the coefficient $\tilde C_0 =(\tilde q-1) / (\tilde q+1)$, the function $\langle \tilde \alpha^2 \rangle^{(\alpha)}_{\tilde K} = \langle \tilde \alpha^2 \rangle^{(\alpha)} \, \tilde E_\chi(\tilde K, \Omega)$ and we use the following model for the spectrum function $\tilde E_\chi(\tilde K, \Omega) = k_\chi^{-1} \, (\tilde q-1) \, (\tilde K / k_{\chi})^{- \tilde q}  \, \delta(\Omega)$, where the wave number $k_{\chi} = 1 / l_{\chi}$ and we take into account that $l_0 \ll l_\chi$.

As follows from Eq.~(\ref{BB15}), the parameter $\sigma_\alpha > 0$. This implies that  homogeneous kinetic helicity fluctuations in a homogeneous turbulence with shear act against mean-field dynamo (see Eqs.~(\ref{BB13}) and~(\ref{E5})).
For small hydrodynamic and magnetic Reynolds numbers (i.e., for the range of validity of SOCA) the parameter $\sigma_{_{B}}$ is negative and the shear-current effect cannot generate the large-scale magnetic field (see \cite{RAS06,RUK06}). This result is in agreement with \cite{RK04} (see also Sec.~V).

\subsection{Inhomogeneous kinetic helicity fluctuations}

Now let us consider inhomogeneous kinetic helicity fluctuations.
The last term in Eq.~(\ref{C14}) determines a large-scale drift velocity of the mean magnetic field:
\begin{eqnarray}
{\bf V}^{(\alpha)} = {1 \over 2} \, \bec{\nabla} \int G_T \, \langle \tilde{\alpha}^2 \rangle^{(\alpha)}_{\tilde K} \, d\tilde K \, d\Omega = {\tilde C_0 \over 2} \, \tilde \tau_\chi \, \bec{\nabla} \langle \tilde{\alpha}^2 \rangle^{(\alpha)} \;,
\nonumber\\
\label{CB1}
\end{eqnarray}
and the third term in Eq.~(\ref{C14}) describes a negative contribution to the turbulent magnetic diffusion of the mean magnetic field \cite{K76} (see also \cite{M78}):
\begin{eqnarray}
\eta^{(\alpha)}_{_{T}} = - \int G_T \, \langle \tilde{\alpha}^2 \rangle^{(\alpha)}_{\tilde K} \, d\tilde K \, d\Omega = - \tilde C_0 \, \tilde \tau_\chi \, \langle \tilde{\alpha}^2 \rangle^{(\alpha)}
\; .
\label{CB2}
\end{eqnarray}
The second term in Eq.~(\ref{C14}) determines the mean $\alpha$ effect caused by a combined action of a large-scale shear and inhomogeneous kinetic helicity fluctuations:
\begin{eqnarray}
\bar{\alpha}^{(S,\alpha)} &=& - {S \over 2} \, {\nabla}_z \int G_T \, \big(G_T + 2\tilde C_2 \, \tau_0\big) \, \langle \tilde \alpha^2 \rangle^{(\alpha)}_{\tilde K} \, d\tilde K \, d\Omega
\nonumber\\
&=& - \sigma_\alpha {\tilde \tau_\chi^2 \, S \over 2} \, {\nabla}_z \langle \tilde{\alpha}^2 \rangle^{(\alpha)} \; .
\label{CB3}
\end{eqnarray}
The first term in Eq.~(\ref{C14}) that is proportional to $\bar J_x$, contributes to the coefficient $\sigma_\alpha$ determined by Eq.~(\ref{BB15}).

Therefore, both approaches (the spectral tau-approximation and SOCA) demonstrate that homogeneous kinetic helicity fluctuations alone with zero mean value in a sheared homogeneous turbulence cannot cause large-scale dynamo. On the other hand, inhomogeneous kinetic helicity fluctuations can generate large-scale magnetic field in a sheared turbulence.

\section{Threshold for shear-current dynamo versus Reynolds number}

In Sections~III and~IV we have shown that homogeneous kinetic helicity fluctuations with zero mean in a sheared turbulence act against mean-field dynamo. On the other hand, shear-current effect can generate large-scale magnetic field even in a homogeneous nonhelical turbulence with large-scale shear. The shear-current dynamo has been studied in \cite{RK03,RK04,RKL06,RKCL06} for large hydrodynamic and magnetic Reynolds numbers. In this Section we demonstrate that hydrodynamic and magnetic Reynolds numbers can be not large in order to maintain the shear-current dynamo. To this end we examine the threshold required for the generation of a large-scale magnetic field by the shear-current dynamo.

Let us neglect the effect of kinetic helicity fluctuations discussed in previous sections (i.e., consider the case when $\langle \tilde \alpha^2 \rangle^{(\alpha)} \ll u_0^2$). A general form of the parameter $\sigma_{_{B}}$ entering in Eqs.~(\ref{BB13}) and~(\ref{E2}) and defining the shear-current effect is
given by
\begin{eqnarray}
\sigma_{_{B}} = {1 \over 15\tau_0^{2}} \, \int \biggl(1 + {k \,(d\tau_r / dk)
\over \tau_r(k)} \biggr) \, \tau_r^2(k) \, E(k) \, dk \; .
 \label{E1}
\end{eqnarray}
Equation~(\ref{E1}) has been derived in \cite{RKL06} using the $\tau$ approach.
Here $E(k)$ is the energy spectrum function, $\tau_r(k)$
is the relaxation time of the cross-helicity
tensor that determines the mean electromotive force.
For large hydrodynamic and magnetic Reynolds numbers
the relaxation time $\tau_r(k)$ of the cross-helicity tensor is of the
order of the correlation time of turbulent velocity field $\tau(k)$.
For simplicity we consider the case $\epsilon=0$.

\subsection{Developed turbulence at low magnetic Prandtl numbers}

\begin{figure}
\vspace*{2mm} \centering
\includegraphics[width=8cm]{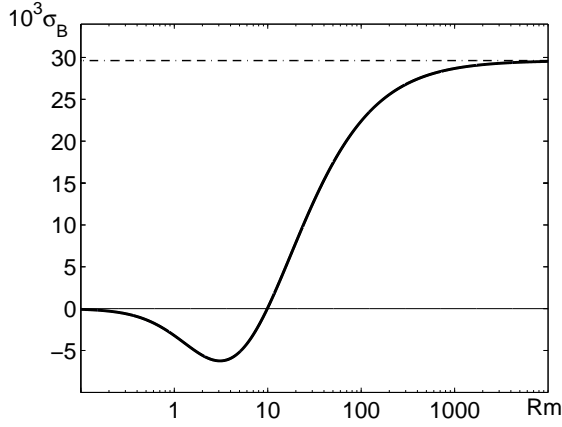}
\caption{\label{Fig1} The coefficient $\sigma_{_{B}}$ defining the
shear-current effect versus the magnetic Reynolds number ${\rm Rm}$
for very large hydrodynamic Reynolds numbers. The dashed-dotted
horizontal line $\sigma_{_{B}}=4/135$ corresponds to the case of
large magnetic Reynolds numbers.}
\end{figure}

Let us first consider a developed turbulence at low magnetic
Prandtl numbers. In this case the relaxation time $\tau_r(k)$ of the cross-helicity tensor which takes into account the magnetic diffusion $\eta$ due to electrical conductivity of the fluid, is determined by
equation:
\begin{eqnarray}
\tau_r^{-1}(k) = \eta \, k^2 + (C \, \tau_0)^{-1} \, \Big({k \over
k_0}\Big)^{\mu} \;,
 \label{AA1}
\end{eqnarray}
where we take into account that $\nu \ll \eta$. For example, the
Kolmogorov scaling corresponds to $\mu=2/3$, i.e., $\tau_r(k) \propto
k^{-2/3}$ (large hydrodynamic and magnetic Reynolds numbers), while
for small magnetic Reynolds numbers, the time $\tau_r(k) \propto 1/
(\eta k^2)$, i.e., $\tau_r(k) \propto k^{-2}$. Using Eqs.~(\ref{E1})
and~(\ref{AA1}) we determine the parameter $\sigma_{_{B}}$ defining
the shear-current effect versus the magnetic Reynolds number:
\begin{eqnarray}
\sigma_{_{B}} &=&  {16 \over 135} \, \biggl[1 + {15 \, \pi \over
\sqrt{2} \, {\rm Rm}^{3/2}} \, \Big(1 - {2 \over \pi} \, \arctan
\sqrt{2/{\rm Rm}}\Big)
\nonumber \\
&& - {6 \over {\rm Rm}} \, \Big(2 + {1 \over 2 + {\rm Rm}} \Big) -
{3 \over 4} \, \Big({{\rm Rm} \over 2 + {\rm Rm}}\Big)^2 \biggr]
\; .
 \label{AA2}
\end{eqnarray}
The asymptotic formulas for
the parameter $\sigma_{_{B}}$ are as follows.
When ${\rm Rm} \ll 1$ the parameter $\sigma_{_{B}}$ reads
\begin{eqnarray*}
\sigma_{_{B}} = - {2  \over 105} \, {\rm Rm}^2 \; .
\end{eqnarray*}
This implies that there is no shear-current dynamo for ${\rm Rm} \ll
1$ in a developed turbulence at low magnetic Prandtl numbers. When
${\rm Re} \gg {\rm Rm} \gg 1$ the parameter $\sigma_{_{B}}$ is
\begin{eqnarray*}
\sigma_{_{B}} = {4 \over 135} \, \Big(1 - {36 \over {\rm Rm}} \Big)
\; .
\end{eqnarray*}
The coefficient $\sigma_{_{B}}$ defining the shear-current effect
versus the magnetic Reynolds number is shown in Fig.~1.
This figure demonstrates that in a developed turbulence at low magnetic
Prandtl numbers the threshold  in the magnetic Reynolds number ${\rm
Rm}_{\rm cr}$ required for the shear-current dynamo is ${\rm Rm}_{\rm cr}
\approx 10$.

\subsection{A random flow with a scale-independent correlation
time}

Let us consider a random flow with a scale-independent
correlation time. In this case the exponent $\mu=0$, the coefficient
$C=1$ and the relaxation time $\tau_r(k)$ of the cross-helicity tensor
which takes into account kinematic viscosity $\nu$ and the magnetic
diffusion $\eta$ due to electrical conductivity of the fluid, is
determined by equation:
\begin{eqnarray}
\tau_r^{-1}(k) = (\nu + \eta) \, k^2 + \tau_0^{-1} \; . \label{TA4}
\end{eqnarray}
In this case the turbulent energy spectrum function is
\begin{eqnarray}
E(k) = k_0^{-1} \, (q-1) \Big[1 - {\rm
Re}^{(1-q)/2} \Big]^{-1} \, \biggl({k \over k_{0}}\biggr)^{-q} \; .
\label{TA5}
\end{eqnarray}
Note that in a random flow with a scale-independent
correlation time the viscous scale is $l_\nu = l_0 / \sqrt{\rm Re}$, while
the viscous scale of the Kolmogorov turbulence is $l_\nu = l_0 / {\rm Re}^{3/4}$. Here the hydrodynamic Reynolds number ${\rm Re} > 1$.

When the exponent $q=0$, the parameter $\sigma_{_{B}}$ defining the shear-current
effect reads
\begin{eqnarray}
\sigma_{_{B}} &=&  {1  \over 60 \, (\sqrt{\rm Re} -1)} \,
\biggl[{\sqrt{\rm Re} \, (3 + a \, {\rm Re}) \over (1 + a \, {\rm
Re})^2} - {3 + a \over (1 + a)^2}
\nonumber \\
&& + {1 \over \sqrt{a}} \, \Big(\arctan\sqrt{a \, {\rm Re}} -
\arctan\sqrt{a} \Big) \biggr] \;,
 \label{A6}
\end{eqnarray}
where $a={\rm Rm}^{-1}+ {\rm Re}^{-1}$. The asymptotic formulas for
the parameter $\sigma_{_{B}}$ are as follows. When ${\rm Rm} \ll 1$ and
${\rm Pr}_m \ll 1$, the parameter $\sigma_{_{B}}$ is
\begin{eqnarray*}
\sigma_{_{B}} = - {{\rm Rm}^2 \over 30 \, {\rm Re}^{3/2}} \,
\big({\rm Re} + \sqrt{\rm Re} + 1\big) \;,
\end{eqnarray*}
where ${\rm Pr}_m = \nu/\eta$ is the magnetic Prandtl number.
For ${\rm Rm} \gg {\rm Re} \gg 1$ the parameter
$\sigma_{_{B}} \approx 1/38$, while for ${\rm Re} \gg {\rm Rm} \gg 1$
the parameter $\sigma_{_{B}} \approx 1/34$.

When $q=2$ the parameter $\sigma_{_{B}}$ defining the shear-current
effect reads
\begin{eqnarray}
\sigma_{_{B}} &=&  {1  \over 15 \, (\sqrt{\rm Re} -1)} \, \biggl\{
\sqrt{\rm Re} \, \biggl[1 + {a \, (7 + 5 \, a) \over 4\, (1 + a)^2}
\nonumber \\
\nonumber \\
&&  + {9 \, \sqrt{a} \over 4} \Big(\arctan\sqrt{a} - \arctan\sqrt{a
\, {\rm Re}} \Big) \biggr]
\nonumber \\
&& - {a \, {\rm Re} \, (7 + 5 \, a\, {\rm Re}) \over 4 \, (1 + a \,
{\rm Re})^2}  -1  \biggr\} \; .
 \label{A8}
\end{eqnarray}
When ${\rm Rm} \ll 1$ and ${\rm Pr}_m \ll 1$, the parameter
$\sigma_{_{B}}$ is
\begin{eqnarray*}
\sigma_{_{B}} = - {3  \over 20 \, (\sqrt{\rm Re} -1)}  \;,
\end{eqnarray*}
while for ${\rm Rm} \gg {\rm Re} \gg 1$ the parameter $\sigma_{_{B}}
\approx 1/15$. For ${\rm Re} \gg {\rm Rm} \gg 1$ the parameter
$\sigma_{_{B}}$ is
\begin{eqnarray*}
\sigma_{_{B}} = {1  \over 15} \, \biggl[1 - {9 \, \pi \, \sqrt{\rm
Re} \over 8 \, {\rm Rm}} \biggr] \; .
\end{eqnarray*}
The coefficient $\sigma_{_{B}}$ defining the shear-current effect
versus the magnetic Reynolds number for a random flow
with a scale-independent correlation time is shown in Fig.~2 (for
$q=0$). The function $\sigma_{_{B}}({\rm Rm})$ for the exponent $q=2$ is similar to that for $q=0$. The threshold  in the magnetic
Reynolds number ${\rm Rm}_{\rm cr}$ required for the shear-current dynamo versus the hydrodynamic Reynolds numbers ${\rm Re}$ is shown in
Fig.~3. Figure~3 demonstrates that the hydrodynamic and magnetic
Reynolds numbers can be of the order of 1 in order to maintain the
shear-current dynamo and there is no need for a developed inertial
range.

\begin{figure}
\vspace*{2mm} \centering
\includegraphics[width=8cm]{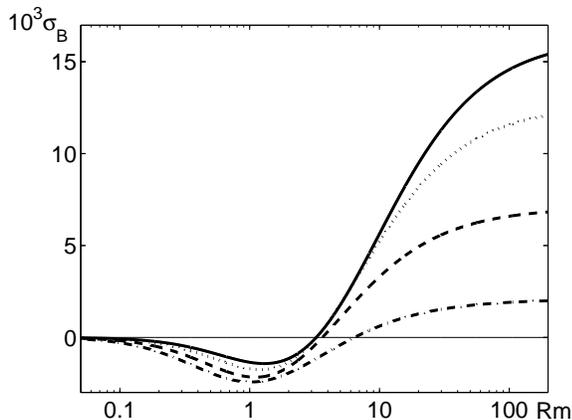}
\caption{\label{Fig2} The coefficient $\sigma_{_{B}}$ defining the
shear-current effect versus the magnetic Reynolds number ${\rm Rm}$
for a random flow with a scale-independent correlation time and
$q=0$. The different curves corresponds to the following values of
the hydrodynamic Reynolds numbers ${\rm Re}$: $\; \; {\rm Re}=10$
(solid); ${\rm Re}=6$ (dotted); ${\rm Re}=3$ (dashed); ${\rm
Re}=1.5$ (dashed-dotted).}
\end{figure}

\begin{figure}
\vspace*{2mm} \centering
\includegraphics[width=8cm]{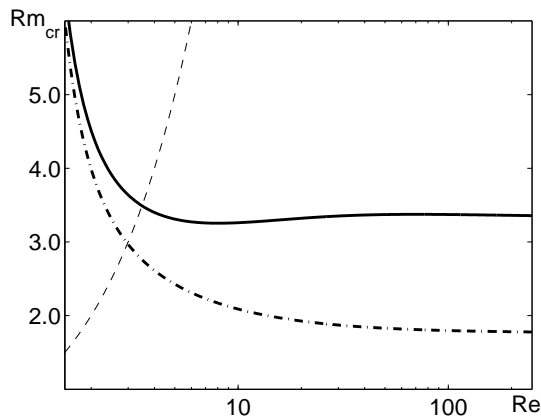}
\caption{\label{Fig3} The threshold in the magnetic Reynolds number
${\rm Rm}_{\rm cr}$ for the shear-current dynamo versus the
hydrodynamic Reynolds numbers ${\rm Re}$ for a random flow with a
scale-independent correlation time and $q=0$ (solid line) and $q=2$
(dashed-dotted line). The dashed line is ${\rm Pr}_m =\nu/\eta=1$.}
\end{figure}

The mean-field dynamo instability due to shear-current effect is saturated by nonlinear effects. A dynamical nonlinearity in the mean-field dynamo which determines the evolution of small-scale magnetic helicity, is of a great importance due to the conservation law for the total (large and small scales) magnetic helicity in turbulence with very large magnetic Reynolds numbers (see, e.g.,
\cite{KR82,KRR94,GD94,KR99,KMRS2000,BF00,VC01,BB02}, and a review
\cite{BS05}). In particular, the mean-field dynamo is essentially nonlinear due to the evolution of the small-scale magnetic helicity (\cite{GD94}). Even for very small mean magnetic field the
magnetic $\alpha$ effect that is related to the small-scale magnetic helicity, is not small.

The nonlinear mean-field dynamo due to a shear-current effect has been studied in \cite{RKL06}, whereby the transport of magnetic helicity as a dynamical nonlinearity has been taken into account. It has been demonstrated in \cite{RKL06} that the magnetic helicity flux strongly affects the magnetic field dynamics in the nonlinear stage of the shear-current dynamo. Numerical solutions of the nonlinear mean-field dynamo equations which take into account the shear-current effect (see \cite{RKL06}), show that if the magnetic helicity flux is not small, the saturated level of the mean magnetic field is of the order of the equipartition field determined by the turbulent kinetic energy. The results of this study are in a good agreement with numerical simulations performed in \cite{B05,BHK05,BS05B}. Note that a non-zero magnetic helicity flux is related to open boundary conditions (see \cite{KMRS2000,BF00}, and a review \cite{BS05}). Finally, we point out an important issue related to the gauge invariance formulation for the magnetic helicity that is described in details in \cite{BS05,SB06}.

\section{Discussion}

We study effects of kinetic helicity fluctuations in a homogeneous turbulence with large-scale shear using the spectral tau-approximation and the second order correlation approximation (SOCA). We show that homogeneous kinetic helicity fluctuations  alone with a zero mean cannot cause a large-scale dynamo in a sheared turbulence. This negative result based on SOCA and tau approach, is a quantitative one: the sign of a certain coefficient in the mean electromotive force $(\propto -\sigma_{\alpha})$ turns out to be unfavorable for the mean-field dynamo. This result is in a contradiction to that suggested in \cite{PROC07} using phenomenological arguments. In order to compare with the results obtained in \cite{PROC07}, we rewrite Eqs.~(\ref{E2}) and~(\ref{E3}) in the following form:
\begin{eqnarray}
{\partial \bar A \over \partial t} &=& \sigma_{_{S}} \, S \, l_0^2 \, \bar B'_y + \tilde \eta_{_{T}} \, \bar A''  \;,
 \label{EE2}\\
{\partial \bar B_y \over \partial t} &=& - S \, \bar A' + \tilde \eta_{_{T}} \, \bar B''_y
\;,
 \label{EE3}
\end{eqnarray}
where the mean magnetic field, $\bar{\bf B} = \bar B_y(z) \, {\bf e}_y + \bec{\nabla} {\bf \times} [\bar A(z) \, {\bf e}_y]$. Equation~(\ref{EE3}) is similar to Eq.~(8) derived in \cite{PROC07}, whereby $\Omega'$ is replaced by $-S$, while Eq.~(\ref{EE2}) is similar to Eq.~(9) derived in \cite{PROC07}.

The first term in the right hand side of Eq.~(\ref{EE2}) determines two different effects, namely the shear-current effect $\propto \sigma_{_{B}} \, S \, \bar B'_y$ and the effect $\propto -\sigma_{\alpha}\, S \, \bar B'_y$ caused by the homogeneous kinetic helicity fluctuations in a sheared turbulence. The shear-current effect  results in the mean-field dynamo, while the second effect $(\propto -\sigma_{\alpha})$ acts against the mean-field dynamo (see Eqs.~(\ref{BB13})-(\ref{B15}) and~(\ref{BB15})). These two effects can be interpreted as the off-diagonal terms in the tensor of turbulent magnetic diffusion, and they cannot be reduced to the standard $\alpha$ effect.

Note that the first term in the right hand side of Eq.~(\ref{EE2}) that determines the effect of homogeneous kinetic helicity fluctuations in a sheared turbulence, is similar to the first term in the right hand side of Eq.~(9) derived in \cite{PROC07} except for it has opposite sign. The latter is crucial for the mean-field dynamo. In this comparison we have not taken into account the shear-current effect determined by the coefficient $\sigma_{_{B}}$, because this effect has not been considered in \cite{PROC07}. The results obtained in the present study are derived using the rigorous mean field theory based on SOCA (see Sec.~IV). These results are also in a good agreement with those obtained by the tau-approach (see Sec.~III). On the other hand, the results of derivation performed in \cite{PROC07} by phenomenological arguments using ad hoc mean-field equations are in a disagreement with our results.

The shear-current effect causes the mean-field dynamo in a homogeneous nonhelical turbulence with imposed large-scale shear. This effect has been studied previously (see \cite{RK03,RK04,RKL06}) only for large hydrodynamic and magnetic Reynolds numbers. In the present study we determine the threshold required for the shear-current dynamo as a function of Reynolds number and demonstrate that threshold value of the Reynolds number is of the order of 1. This implies that there is no need for a developed inertial range in order to maintain the shear-current dynamo.

In the present study we also recover the results obtained in \cite{K76} (see also \cite{M78}) for a shear-free turbulence, whereby a negative contribution of kinetic helicity fluctuations to the turbulent magnetic diffusion, $\eta^{(\alpha)}_{_{T}} = - \tau_\chi \, \langle \tilde{\alpha}^2 \rangle^{(\alpha)}$, and a large-scale drift velocity of the mean magnetic field, ${\bf V}^{(\alpha)} \propto \tau_\chi \, \bec{\nabla} \langle \tilde{\alpha}^2 \rangle^{(\alpha)}$, have been found.

On the other hand, we have demonstrated that inhomogeneous kinetic helicity fluctuations in a sheared turbulence cause a nonzero mean alpha effect, $\bar{\alpha}^{(S,\alpha)} \propto - \tau_\chi^2 \, S \, {\nabla}_z \langle \tilde{\alpha}^2 \rangle^{(\alpha)}$, where the mean vorticity due to the large-scale shear is ${\bf W}^{(S)} = S \, {\bf e}_z$. The mean alpha effect $\bar{\alpha}^{(S,\alpha)}$ is formed by a combined action of a large-scale shear in turbulent flow and inhomogeneous kinetic helicity fluctuations even when $\langle \tilde{\alpha} \rangle^{(\alpha)}=0$. The large-scale shear and the mean alpha effect can cause a mean-field dynamo (see \cite{SIL2000}) that is similar to the $\alpha\Omega$-dynamo.

The discussed effects in this study might be important in astrophysics
(e.g., accretion discs, colliding protogalactic clouds, merging protostellar clouds \cite{RKCL06}) and laboratory dynamo experiments. In particular, non-symmetrical explosions of supernova may produce fluctuations of kinetic helicity located in larger scales than  small-scale turbulence existing in convective zones inside stars. On the other hand, the shear-current dynamo acts together with the $\alpha$-shear dynamo which is similar to the $\alpha \Omega$ dynamo. The shear-current effect does not quenched (see \cite{RK04,RKL06}) contrary to the quenching of the nonlinear $\alpha$ effect, the turbulent magnetic diffusion, the effective drift velocity. This implies that the shear-current effect might be the only surviving effect, which can explain the origin of large-scale magnetic fields in sheared astrophysical turbulence.

\begin{acknowledgments}
We have benefited from stimulating discussions with Alexander Schekochihin and Dmitry Sokoloff.
\end{acknowledgments}

\appendix

\section{Derivation of Eqs.~(\ref{B3}) and~(\ref{B11}) using the
$\tau$-approach}

In order to study the effect of kinetic helicity fluctuations in a
sheared turbulence we use a procedure applied in \cite{RK04} for a
sheared turbulence. Let us derive equations for the second moments. To
exclude the pressure term from the equation of motion~(\ref{A1}) we
calculate $\bec{\nabla} {\bf \times} (\bec{\nabla} {\bf \times} {\bf
u})$. Then we rewrite the obtained equation and
Eq.~(\ref{A2}) in a Fourier space. We also apply the
two-scale approach, e.g., we use large scale ${\bf R} = ( {\bf x} +
{\bf y}) / 2$, $\, {\bf K} = {\bf k}_1 + {\bf k}_2$ and small scale
${\bf r} = {\bf x} - {\bf y}$, $\, {\bf k} = ({\bf k}_1 - {\bf k}_2)
/ 2$ variables (see, e.g., \cite{RS75}). We derive
equations for the following correlation functions:
\begin{eqnarray}
f_{ij}({\bf k}) &=& \hat L(u_i; u_j) \;, \quad h_{ij}({\bf k}) =
\hat L(b_i; b_j) \;,
\nonumber\\
g_{ij}({\bf k}) &=& \hat L(b_i; u_j) \;,
\label{B1}
\end{eqnarray}
where
\begin{eqnarray*}
\hat L(a; c) = \int \langle a({\bf k} + {\bf  K} / 2) c(-{\bf k} +
{\bf  K} / 2) \rangle
\\
\times \exp{(i {\bf K} {\bf \cdot} {\bf R}) } \,d {\bf  K} \;,
\end{eqnarray*}
where $\langle ... \rangle$ denotes averaging over ensemble
of turbulent velocity field. The equations for these correlation
functions are given by
\begin{eqnarray}
{\partial f_{ij}({\bf k}) \over \partial t} &=& i({\bf k} {\bf
\cdot} {\bf B}) \Phi_{ij} + I^f_{ij} + I_{ijmn}^S({\bf U})
f_{mn}
\nonumber \\
&& + \hat{\cal N} f_{ij} + F_{ij}\;,
\label{B6} \\
{\partial h_{ij}({\bf k}) \over \partial t} &=& - i({\bf k}{\bf
\cdot} {\bf B}) \Phi_{ij} + I^h_{ij} + E_{ijmn}^S({\bf U})
h_{mn} + \hat{\cal N} h_{ij} \;,
\nonumber \\
\label{B7} \\
{\partial g_{ij}({\bf k }) \over \partial t} &=& i({\bf k} {\bf
\cdot} {\bf B}) [f_{ij}({\bf k}) - h_{ij}({\bf k}) - h_{ij}^{(H)}] +
I^g_{ij}
\nonumber \\
&& + J_{ijmn}^S({\bf U}) g_{mn} + \hat{\cal N} g_{ij} \;,
\label{B8}
\end{eqnarray}
(see \cite{RK04}), where hereafter we omit argument $t$ and ${\bf R}$ in the
correlation functions and neglect small terms $ \sim O(\nabla^2)$.
Here $F_{ij}$ is related to the forcing terms and $\bec{\nabla} = \partial / \partial {\bf R} $. In Eqs.~(\ref{B6})-(\ref{B8}), $\Phi_{ij}({\bf k}) = g_{ij}({\bf k}) -
g_{ji}(-{\bf k})$, and $ \hat{\cal N} f_{ij}$, $\, \hat{\cal N}h_{ij}$, $\, \hat{\cal N}g_{ij}$, are the third-order moment terms appearing due to the nonlinear terms which include also molecular dissipation terms. The tensors $I_{ijmn}^S({\bf U})$, $\, E_{ijmn}^S({\bf U})$ and
$J_{ijmn}^S({\bf U})$ are given
by
\begin{eqnarray*}
I_{ijmn}^S({\bf U}) &=& \biggl[2 k_{iq} \delta_{mp} \delta_{jn}
+ 2 k_{jq} \delta_{im} \delta_{pn} - \delta_{im} \delta_{jq}
\delta_{pn}
\nonumber\\
&& - \delta_{iq} \delta_{jn} \delta_{pm} + \delta_{im} \delta_{jn}
k_{q} {\partial \over \partial k_{p}} \biggr] \nabla_{p} U_{q} \;,
\nonumber\\
E_{ijmn}^S({\bf U}) &=& \biggl[\delta_{im} \delta_{jq}
\delta_{pn} + \delta_{jm} \delta_{iq} \delta_{pn}
\nonumber\\
& & + \delta_{im} \delta_{jn} k_{q} {\partial \over \partial k_{p}}
\biggr] \nabla_{p} U_{q} \;,
\nonumber\\
J_{ijmn}^S({\bf U}) &=& \biggl[2 k_{jq} \delta_{im} \delta_{pn}
- \delta_{im} \delta_{pn} \delta_{jq} + \delta_{jn} \delta_{pm}
\delta_{iq}
\nonumber\\
& & + \delta_{im} \delta_{jn} k_{q} {\partial \over \partial k_{p}}
\biggr] \nabla_{p} U_{q} \;,
\end{eqnarray*}
where $ k_{ij} = k_i k_j / k^2 $. The source terms $I_{ij}^f$ , $\,
I_{ij}^h$ and $I_{ij}^g$ which contain
the large-scale spatial derivatives of the magnetic field ${\bf B}$, are
given in \cite{RK04}. Next, in Eqs.~(\ref{B6})-(\ref{B8}) we split the tensor for
magnetic fluctuations into nonhelical, $h_{ij},$ and helical,
$h_{ij}^{(H)},$ parts. The helical part of the tensor of magnetic
fluctuations $h_{ij}^{(H)}$ depends on the magnetic helicity and it
follows from the magnetic helicity conservation arguments (see, {\rm
e.g.,} \cite{KR82,KRR94,GD94,KR99,KMRS2000,BB02}). We also use the
spectral $\tau$ approximation which postulates that the deviations
of the third-moment terms, $\hat{\cal N} M^{(III)}({\bf k})$, from
the contributions to these terms afforded by the background
turbulence, $\hat{\cal N} M^{(III,0)}({\bf k})$, are
expressed through the similar deviations of the second moments,
$M^{(II)}({\bf k}) - M^{(II,0)}({\bf k})$ [see Eq.~(\ref{A3})].

We take into account that the characteristic time of variation of
the magnetic field ${\bf B}$ is substantially larger than the
correlation time $\tau(k)$ for all turbulence scales. This allows us
to get a stationary solution for Eqs.~(\ref{B6})-(\ref{B8}) for the
second-order moments, $M^{(II)}({\bf k})$, which are the sum of
contributions caused by a shear-free turbulence, a
sheared turbulence and kinetic helicity fluctuations. The contributions to the mean
electromotive force caused by a shear-free turbulence and the sheared turbulence without kinetic helicity fluctuations are
given in \cite{RK04}. On the other hand, the contributions to the
electromotive force caused by a combined action of the sheared turbulence and the kinetic helicity fluctuations, are given by ${\cal E}_{m}^{(S,\alpha)} = \varepsilon_{mji} \, \int \, g_{ij}^{(S,\alpha)}({\bf k}) \,d {\bf k} $,
where the corresponding contributions to the cross-helicity
tensor $g_{ij}^{(S,\alpha)}$ in the kinematic approximation, are given by
\begin{eqnarray}
g_{ij}^{(S,\alpha)}({\bf k}) = i \tau \, \Big[J_{ijmn}^S \, \tau \,    ({\bf k} {\bf \cdot} {\bf B}) + \tau \, ({\bf k} {\bf \cdot} {\bf B}) \, I_{ijmn}^S \Big] \, f_{mn}^{(0,\chi)} \;,
\nonumber\\
\label{B2}
\end{eqnarray}
where $f_{ij}^{(0,\chi)} \equiv \langle u_i \, u_j \rangle^{(0, \chi)} = - i \, \chi^v \, \varepsilon_{ijn} \, k_n \, E_\chi(k)/(8 \pi k^4)$ is the helical part of the velocity fluctuations.  Straightforward calculations using Eq.~(\ref{B2}) yields the contributions $\alpha_{ij}^{(S,\alpha)}$ to the $\alpha$ tensor caused by a combined action of the sheared turbulence and the kinetic helicity fluctuations, that is determined by Eq.~(\ref{B3}). The total $\alpha$ tensor is given by $\alpha_{ij} = \tilde \alpha \, \delta_{ij} + \alpha_{ij}^{(S,\alpha)}$, where $\tilde \alpha \, \delta_{ij}$ determines a shear-free turbulence contribution. This procedure allows us to derive the equation for the evolution of the magnetic field ${\bf B}$ that is given by Eq.~(\ref{B5}).

Using Eq.~(\ref{B5}) we derive equation for the correlation function$\langle \alpha_{ij} B_p \rangle^{(\alpha)}$ that is given by Eq.~(\ref{B9}),
where $\langle ... \rangle^{(\alpha)}$ denote an averaging over random $\alpha$ fluctuations. Equation~(\ref{B9}) include the third-order moments caused by the nonlinear terms. In Eq.~(\ref{B9}) we use the spectral $\tau$ approximation~(\ref{A3}), where the large-scale relaxation time is of the order of $\tau_\chi$. We also take into account that the characteristic time of variation of the mean magnetic field $\bar{\bf B}=\langle {\bf B} \rangle^{(\alpha)}$ is substantially larger than the relaxation time $\tau_\chi$. This yields the correlation function $\langle \alpha_{ij} B_j \rangle^{(\alpha)}$:
\begin{eqnarray}
&& \langle \alpha_{ij} B_j \rangle^{(\alpha)} = {\tau_\chi \over 2} \, \biggl\{2 \varepsilon_{jkm} \, \Big[\bar B_n \, \Big(\delta_{ij} \, \langle \tilde \alpha \, \nabla_k \, \alpha_{mn}^{(S,\alpha)} \rangle^{(\alpha)}
\nonumber\\
&& \quad + \delta_{mn} \, \langle \alpha_{ij}^{(S,\alpha)} \, \nabla_k \, \tilde \alpha \rangle^{(\alpha)}  \Big) + \big(\nabla_k \, \bar B_n\big) \, \Big(\delta_{ij} \, \langle \tilde \alpha \, \alpha_{mn}^{(S,\alpha)} \rangle^{(\alpha)}
\nonumber\\
&& \quad + \delta_{mn} \, \langle \alpha_{ij}^{(S,\alpha)} \, \tilde \alpha \rangle^{(\alpha)}  \Big) \Big] + \big[\delta_{ij} +  \tau_\chi \, \big(\nabla_{j} U_{i}^{(S)}\big) \big] \, \big[2 \bar J_j
\nonumber\\
&&  \quad - (\bar{\bf B} {\bf \times}\bec{\nabla})_j \big]  \, \langle \tilde \alpha^2 \rangle^{(\alpha)} \biggr\} \;,
 \label{B10}
\end{eqnarray}
where $\bar {\bf J}= \bec{\nabla} {\bf \times} \bar {\bf B}$. We consider for simplicity a linear mean velocity shear ${\bf U}^{(S)} = (0,Sx, 0)$ and $ {\bf W}^{(S)} = (0,0,S)$, where $S \, \tau_0 \ll 1$. The mean magnetic field $\bar{\bf B}$ in a most simple form is $\bar{\bf B} = (\bar B_x(z), \bar B_y(z), 0)$. Equation~(\ref{B10}) allows us to determine the correlation function $\langle \alpha_{yj} B_j \rangle^{(\alpha)}$ that is given by Eq.~(\ref{B11}).

\section{Derivation of Eq.~(\ref{C14}) using the SOCA-approach}

In order to study the effect of kinetic helicity fluctuations in a
turbulence with large-scale shear we use a second order correlation
approximation (SOCA) applied in \cite{RAS06} for a sheared turbulence. To exclude the pressure term from the equation of motion~(\ref{A1}) we
calculate $\bec{\nabla} {\bf \times} (\bec{\nabla} {\bf \times} {\bf
u})$, then we rewrite the obtained equation and Eq.~(\ref{A2})
in a Fourier space, apply the two-scale approach (i.e., we use
large-scale and small-scale variables), and we
neglect nonlinear terms in Eqs.~(\ref{A1}) and ~(\ref{A2}).
On the other hand, we keep molecular dissipative terms in these equations. We seek for a solution for fluctuations of velocity and magnetic fields as an expansion for weak velocity shear:
\begin{eqnarray}
{\bf u} &=& {\bf u}^{(0)} + {\bf u}^{(1)} + ... \;,
\label{C4} \\
{\bf b} &=& {\bf b}^{(0)} + {\bf b}^{(1)} + ... \;,
\label{C5}
\end{eqnarray}
where
\begin{eqnarray}
b_i^{(0)}({\bf k}, \omega) &=& G_\eta(k, \omega) \, \biggl[i({\bf k} {\bf
\cdot} {\bf B}) \delta_{ij} - \Big(\delta_{ij} \, k_{m} {\partial \over \partial k_{n}}
\nonumber\\
&& + \delta_{im} \delta_{jn}\Big) (\nabla_{n} B_{m}) \biggr] \, u_j^{(0)}({\bf k}, \omega) \;,
 \label{C1}\\
u_i^{(1)}({\bf k}, \omega) &=& G_\nu(k, \omega) \, \biggl[
2 k_{iq} \delta_{jp} + \delta_{ij} \, k_{q} {\partial \over \partial k_{p}}
\nonumber\\
&& - \delta_{iq} \delta_{jp} \biggr] (\nabla_{p} U_{q}) \, u_j^{(0)}({\bf k}, \omega) \;,
\label{C2}\\
b_i^{(1)}({\bf k}, \omega) &=& G_\eta(k, \omega) \, \biggl\{ \biggl[i({\bf k} {\bf \cdot} {\bf B}) \delta_{ij} - \Big(\delta_{ij} \, k_{m} {\partial \over \partial k_{n}}
\nonumber\\
&& + \delta_{im} \delta_{jn} \Big) \, (\nabla_{n} B_{m}) \biggr] \,     u_j^{(1)}({\bf k}, \omega)
\nonumber\\
&&  + \biggl[\delta_{ij} \, k_{q} {\partial \over \partial k_{p}}
+ \delta_{iq} \delta_{jp}\biggr] (\nabla_{p} U_{q}) \, b_j^{(0)}({\bf k}, \omega) \biggr\} ,
\nonumber\\
\label{C3}
\end{eqnarray}
(for details see \cite{RAS06}). Here $G_\nu(k, \omega) = (\nu k^2 - i \omega)^{-1}$ and $G_\eta(k, \omega) = (\eta k^2 - i \omega)^{-1}$.

Equations~(\ref{C1})-(\ref{C3}) allow us to determine the cross-helicity tensor $g_{ij}^{(1)} = (\langle b_i^{(1)} \, u_j^{(0)} \rangle + \langle b_i^{(0)} \, u_j^{(1)} \rangle + \langle u_j^{(0)} \, b_i^{(1)}  \rangle + \langle u_j^{(1)} \, b_i^{(0)} \rangle) / 2$. This procedure yields the contributions, ${\cal E}_{m}^{(S,\alpha)} = \varepsilon_{mji} \, \int \, g_{ij}^{(1,S,\alpha)}({\bf k}, \omega) \,d {\bf k} \,d \omega$, to the electromotive force caused by a combined action of the sheared turbulence and the kinetic helicity fluctuations. In particular, the $\alpha$ tensor caused by the kinetic helicity fluctuations in a sheared turbulence reads:
\begin{eqnarray}
\alpha_{ij}^{(S,\alpha)} = D_1 \, (\partial U)_{ij}^{(S)} + D_2 \, \varepsilon_{ijn} \, W_n^{(S)} \;,
\label{C6}
\end{eqnarray}
where
\begin{eqnarray}
D_1 &=& {1 \over 60} \, \int \Big(G_\eta \big(11 G_\nu + 11 G_\nu^* + 5 G_\eta\big) + G_\eta^* \big(G_\nu 
\nonumber\\
&& + 11 G_\nu^*
+ 5 G_\eta^*\big) + 4 k\big\{G_\eta G'_\eta
+ \big[G_\eta (G_\nu + G^*_\nu) 
\nonumber\\
&& + G_\eta^* G_\nu^* \big]'
+ G_\eta^* (G_\nu + G^*_\eta)' \big\} \Big) \, \tilde E_\chi(k, \omega) \, dk \, d\omega \;,
\nonumber\\
\label{C7}\\
D_2 &=& {1 \over 24} \, \int \big[G_\eta \big(G_\eta - G_\nu - G_\nu^*\big) 
+ G_\eta^* \big(G_\nu 
\nonumber\\
&& - G_\nu^*\big) \big]\,
\tilde E_\chi(k, \omega) \, dk \, d\omega \; .
\label{C8}
\end{eqnarray}
Here $E(k, \omega) = \tilde E(k, \omega) / 8 \pi k^2$ and we integrate over the angles in ${\bf k}$ space. The total $\alpha$ tensor is given by $\alpha_{ij} = \tilde \alpha \, \delta_{ij} + \alpha_{ij}^{(S,\alpha)}$, where $\tilde \alpha \, \delta_{ij}$ determines a shear-free turbulence contribution. Let us consider the following model for the spectrum function $\tilde E_\chi(k, \omega) = \tilde C_\chi \, k_0^{-1} \, (k / k_0)^{- q} \, \delta(\omega)$. Then integration over $k$ and $\omega$ in Eqs.~(\ref{C7}) and~(\ref{C8}) yields the contributions to the $\alpha$ tensor caused by a combined action of the sheared turbulence and the kinetic helicity fluctuations. In particular, the tensor $\alpha_{ij}^{(S,\alpha)}$ is given by Eq.~(\ref{BB3}).

In the scales $l_0 \ll l \ll l_\chi$ there are fluctuations of kinetic helicity  $\chi^v$ (or fluctuations of $\tilde \alpha$). Now we determine the correlation function $\langle \alpha_{ij} B_p \rangle^{(\alpha)}$. To this end we use Eq.~(\ref{B5}) in which $\eta_{_{T}}$ is replaced by $\eta + \eta_{_{T}}$ and the nonlinear terms are neglected. Solving this equation in a Fourier space we determine the magnetic field $B_j(\tilde{\bf K}, \Omega)$, where the wave number $\tilde{K}$ and the frequency $\Omega$ are in the spatial scales $l_0 \ll l \ll l_\chi$ and the time scales $\tau_0 \ll \tau \ll \tau_\chi$. Multiplying the magnetic field $B_j(\tilde{\bf K}, \Omega)$ by the tensor $\alpha_{ij}= \tilde \alpha \, \delta_{ij} + \alpha_{ij}^{(S,\alpha)}$ and averaging over kinetic helicity fluctuations we determine the correlation function $\langle \alpha_{ij} B_j \rangle^{(\alpha)}$:
\begin{eqnarray}
&& \langle \alpha_{ij} B_j \rangle^{(\alpha)} = {1 \over 2} \int G_T \, \biggl\{2 \varepsilon_{jkm} \, \Big[\bar B_n \, \Big(\delta_{ij} \, \langle \tilde \alpha \, \nabla_k \, \alpha_{mn}^{(S,\alpha)} \rangle^{(\alpha)}
\nonumber\\
&& + \delta_{mn} \, \langle \alpha_{ij}^{(S,\alpha)} \, \nabla_k \, \tilde \alpha \rangle^{(\alpha)}  \Big) + \big(\nabla_k \, \bar B_n\big) \, \Big(\delta_{ij} \, \langle \tilde \alpha \, \alpha_{mn}^{(S,\alpha)} \rangle^{(\alpha)}
\nonumber\\
&& + \delta_{mn} \, \langle \alpha_{ij}^{(S,\alpha)} \, \tilde \alpha \rangle^{(\alpha)}  \Big) \Big] + \biggl[\delta_{ij} +  \big(\nabla_{p} U_{q}^{(S)}\big) \, \biggl(\delta_{jp} \delta_{iq}
\nonumber\\
&& + \delta_{ij} \, \tilde K_{q} {\partial \over \partial \tilde K_{p}} \biggl) \, G_T \biggr]\, \big[2 \bar J_j - (\bar{\bf B} {\bf \times}\bec{\nabla})_j \big]  \, \langle \tilde \alpha^2 \rangle^{(\alpha)}_{\tilde K} \biggr\}\, d\tilde K \, d\Omega \;,
\nonumber\\
 \label{C13}
\end{eqnarray}
where $G_T(\tilde K, \Omega) = (\eta_{_{T}} \tilde K^2 - i \Omega)^{-1}$. We consider a linear mean velocity shear ${\bf U}^{(S)} = (0,Sx, 0)$ and assume that the mean magnetic field $\bar{\bf B}$ has a form: $\bar{\bf B} = (\bar B_x(z), \bar B_y(z), 0)$. This allows us to simplify Eq.~(\ref{C13}) and to determine the correlation function $\langle \alpha_{yj} B_j \rangle^{(\alpha)}$ (see Eq.~(\ref{C14})).

\end{document}